# Laser nano-fabrication inside silicon with spatial beam modulation and non-local seeding

Rana Asgari Sabet[1,2], Aqiq Ishraq[2], Alperen Saltik[1], Onur Tokel[1,2,†]

[1] Department of Physics, Bilkent University, Ankara, 06800, Turkey

[2] UNAM – National Nanotechnology Research Center and Institute of Materials Science and Nanotechnology, Bilkent University, Ankara, 06800, Turkey

[†]Corresponding author. E-mail: otokel@bilkent.edu.tr (O.T.)

**Nano-fabrication in silicon, arguably the most important material for modern technology, has been limited exclusively to its surface. Existing lithographic methods cannot penetrate the wafer surface without altering it[1-3], whereas emerging laser-based subsurface or in-chip fabrication remains at > 1 μm resolution[4]. In addition, available methods do not allow positioning or modulation with sub-micron precision deep inside the wafer[4-8]. The fundamental difficulty of breaking these dimensional barriers are two-fold[4], *i.e.*, complex nonlinear effects inside the wafer and the inherent diffraction limit for laser light[9,10]. Here, we overcome these challenges by using structured-laser-beams and exploiting preformed subsurface structures as seed, in order to establish the first controlled nano-fabrication capability inside silicon. We demonstrate buried nano-structures of feature sizes down to 100 nm ± 25 nm, with sub-wavelength and multi-dimensional control; thereby improving the state-of-the-art by an order-of-magnitude. In order to showcase the photonic capabilities, we fabricated the first in-chip nano-photonic elements, *i.e.,* nano-gratings with record diffraction efficiency and spectral control. The reported advance is an important step towards in-chip nano-photonic systems, micro-/nano-fluidics & NEMS/MEMS, and 3D electronic-photonic integration.**

Silicon is unique for its use as the substrate material for the electronics industry[11], as well as its prominent role in micro-/nano-photonics[11-15]. The material has diverse applications in the near- and mid-infrared regime[16,17]. Nano-lithography techniques are enabling unique Si platforms for studying new physics and exciting functionalities, including sub-wavelength and meta-material technologies[14,16-19]. The plethora of functionality is virtually entirely limited to the wafer surface, and thus may be identified as "on-chip". A distinct paradigm is to directly fabricate devices inside the bulk of Si, without altering the top or bottom surfaces of the wafer. The corresponding "in-chip" fabrication paradigm[4] is based on exploiting infrared lasers where the wafer is transparent, in order to access the bulk of Si. This has already lead to significant advances[20], in particular with the breakthrough of introducing functionality directly and inside Si, *e.g.*, waveguides, lenses, information storage and state-of-the-art diffractive optical elements[4-6,21].

While offering unique photonic capabilities, current applications are constrained by the >1-μm fabrication resolution limit[4]. Overcoming this barrier and achieving multi-dimensional control inside the wafer would be a major advance. It has the potential to enable truly 3D photonics[22-24], introduce unique functionalities beyond conventional optics[25,26], and even lead to nano-photonics and metamaterials inside Si wafers. These would have the added benefit of keeping the chip surface unaltered for additional micro-/nano-functionality. Here, we address this critical fabrication challenge and achieve controlled laser nano-fabrication deep inside the wafer, which in turn enables the first nano-photonic devices inside silicon.

Most volumetric laser-writing relies on Gaussian beams. However, the throughput is generally limited by point-by-point assembly, an inherent limitation of top-down fabrication methods[27]. This was overcome with 3D nonlinear laser lithography, where emergent feedback dynamics[28] arising from the interaction of the laser and semiconductor is exploited[4] for rapid bottom-up self-assembly, but with 1-μm resolution[4]. We consider exploiting non-diffractive beams as a promising direction to simultaneously preserve the high throughput and also extend the resolution to the nano-regime. Previous efforts in this direction have not been successful, due to the challenges in balancing competing requirements[20], *i.e.*, low intensity to preserve the wafer surface, and high pulse energy needed to modify the crystal within its bulk[9,10]. Strong beam delocalization of femtosecond (fs) pulses is a further challenge[9,10]. We overcome these using non-diffracting nanosecond (ns) laser pulses in conjunction with a novel nonlocal seeding effect (Fig. 1). The non-diffracting nature of beam propagation and inherent nonlinear thresholding enables one-dimensional (1D) confinement at nanoscale deep inside Si (Fig. 1a), with unscathed wafer surfaces. The strong energy confinement of the spatially-modulated laser pulse allows even beyond-diffraction-limit subsurface nano-patterning (Fig. 1a). We further introduce a second laser-writing modality, where nonlocal seeding from preformed structures allows 2D-nano-confinement and polarisation-controlled nano-lithography (Fig. 1b).

We start by discussing the modulated beam profile allowing such unique features (Fig. 2a). We use Gaussian output of a custom-built fibre laser operating with 10-ns pulses, with up to 9 W power, at the wavelength of λ = 1.55 μm where Si is transparent (see Methods). In order to generate the required strong energy confinement inside the material, the laser is modulated with a spatial light modulator (SLM), imprinting an axicon phase of zeroth-order Bessel function of the first kind on the beam. The Bessel zone length ($z_B$) and core diameter ($d_B$) after SLM are related as, $z_B = w_0/\tan(\theta)$ and $d_B = 2.4/k.\tan(\theta)$, where $w_0$ is radius of the incident beam, $k$ is the wave vector, and $\theta$ is the cone angle of virtual axicon. A compromise arises here, where smaller core diameters also correspond to smaller Bessel zone lengths, potentially impeding high-aspect-ratio nano-lithography. This can be overcome by implementing a lens-axicon doublet, where strong focusing is invoked by adding an aspheric lens ($L_3$ in Fig. 2a). In this manner, $d_B$ and $z_B$ are decoupled; achieving high-aspect-ratio subsurface nano-lithography with powerful scaling and depth control based on laser pulse energy (see Supplementary Note 1).

In order to generate Bessel beams with SLM, we implement virtual phase profiles $\phi(r)$ of the form,

$$\phi(r) = e^{\pm\left(i2\pi\frac{r}{r_0}\right)}, \tag{2}$$

where $r$ is radial position and $r_0$ parameter is related to the cone angle $\theta$ as $r_0 = \lambda/\tan(\theta)$. A positive (negative) sign in the complex argument corresponds to a diverging (converging) axicon. The use of negative sign could at first seem more intuitive, since it would not require the use of $L_3$, in contrast to the positive sign case. However, we did not observe well-controlled in-chip micro-/nano-lithography with the negative sign, corroborating previous work[9]. Thus, diverging axicon-lens doublet is exploited for nano-lithography, integrated to the 4-$f$ system for relaying, scaling and spatial filtering the laser (Fig. 2a).

First, we create Bessel beam profiles in air, to be evaluated experimentally and numerically. For optical characterisation, focused beam is imaged after $L_3$ and its intensity is recorded with 2-μm increments along the optical axis, using an InGaAs camera (Artray, Artcam-03ITNIR) coupled to a magnification system. Fig. 2b-ii shows the experimental 2D intensity profile of a representative Bessel beam ($r_0 = 10$). We represent $r_0$ in units of pixels, i.e., $r_0 = 10$ pixels corresponds to $r_0 = 10 \times 20$ μm. The origin of abscissa in Fig. 2b-ii corresponds to the focal point of $L_3$; followed by the onset of Bessel zone 100 μm later. During lithography, this range is further scaled inside Si due to the high air-Si index contrast. The transverse profile at the maximum intensity point along the optical axis is compared with simulations (see Methods for details), and shown in Fig. 2b-i along a line passing through the centre of camera image (Fig. 2b-i, inset). The comparison confirms the creation of zeroth-order Bessel function at the output of the SLM-lens doublet in air, which is preserved for a wide range of $r_0$ parameters. Thus, simply by digitally tuning the $r_0$ parameter, modulated beams with diverse energy confinement can be created, analogous to creating diverse spatial profiles using different physical optics. We exploit such tunable beams of "zeroth-order Bessel function of the first kind" for nano-lithography, enabling tunable lithographic feature sizes.

In order to exert control at nanoscale inside Si, we focus the Bessel beam directly inside the chip, invoking a nonlinear response[4,20] along the non-diffracting, high-intensity Bessel zone. This in turn induces confined energy deposition and permanent material change. It is important to note that wafer surfaces, as well as bulk crystal above and below lithographic patterns are unaltered after the process. By scanning the sample perpendicular to laser propagation direction, we create planes of one-dimensional (1D) nano-confinement (Fig. 2c-i) deep inside the bulk, with adjustable thickness along $x$-axis and strong depth control. The SEM image of the first laser-written nano-pattern (inset, Fig. 2c-i), indicates 600 nm feature size for an individual nano-plane, and uniform elongation along $z$ axis of 200 μm. This is achieved with $r_0 = 10$, using a laser pulse of 4-μJ energy and circular polarisation, and single scan per nano-plane (Fig. 2c-ii, left). A detailed analysis of influence of polarisation and pulse energy will be given later.

Next, we introduce laser lithography of subsurface structures with 2D nano-confinement (*i.e.*, nano-lines). Direct application of the preceding approach fails to produce nano-lines. Towards this goal, one can resort to longitudinal scanning (along laser propagation axis) or simply exposing Si to laser without any scanning, however; extensive experiments reveal only micro-patterns for a wide range of focusing, laser and SLM parameters (see Supplementary Note 2). We overcome this challenge by invoking a previously unobserved effect, schematically described in Fig. 1b. We first create a nano-plane in Si using the transverse writing modality (Fig. 2c-ii, left). This structure or preform will act as seed in order to create consecutive 2D-confined nano-patterns (Fig. 2c-ii, right & Fig. 2d-i). In order to reveal buried nano-lines, the sample is polished from the $x$-$y$ plane; followed by treatment with selective etchant developed before[4]. The existence of preform is required, as confirmed by Fig. 2d-ii, where subsurface nano-lines simply do not form when the preform is omitted. Further, the seed is required to be within ~3 μm distance in order to induce the formation of a nano-line (see Supplementary Notes 2 & 3). Once the nano-line formation is initiated, it can be elongated across the entire wafer, simply by scanning the wafer along the optical axis. This approach also enables the creation of nano-arrays at all depths where preform is absent.

A representative nano-line from a larger array is shown in Fig. 2d-iii at its $x$-$y$ cross-section (the colour coded subsurface plane is indicated in Fig. 2c-ii, right). The associated preform is created with $r_0 = 10$, while the 300-nm diameter nano-line is created with $r_0 = 6$, laser pulse energy of 6 μJ, repetition rate 150 kHz, circular polarisation, and scanning speed of 3 mm/s. The chemical etching is performed for 40 seconds to reveal the 300-nm cross-section (see Methods). Such structures can be arranged in arbitrary arrays (Fig. 1b, Fig. 2d-i & Fig. 3), potentially covering the entire wafer, which would be limited only by the translation stage range. Thus, the method introduces a powerful in-chip nano-fabrication capability. A model based on spatial intensity profile and thresholding is provided in Supplementary Note 3, explaining our empirical observations on emergence and nano-patterning.

Until now, we focused on the emergence and propagation of in-chip nano-structures, and constrained the discussion to circular polarisation for maximum symmetry. In order to further push the limits of nano-lithography and to create diverse nano-arrays, we inquire the effect of laser polarisation (Fig. 3). First we inquire any polarisation effects for 1D-confined structures. We observe that when polarisation is set to linear and parallel to scanning direction, the feature size is reduced (see Supplementary Note 4). Using this geometry, systematic experiments were performed for phase ($r_0$) and pulse energy ($E_p$). We find that by simultaneously decreasing $r_0$ and $E_p$, one can reduce the feature size to an order-of-magnitude smaller than state-of-the-art, down to 100 nm (Fig. 3a). A representative SEM image given in Fig. 3c shows uniform nano-structuring with $r_0 = 10$, achieving a diffraction-limited feature size of 250 nm.

A remarkable phenomenon is observed for in-chip lithography, when one considers 2D-confined nano-patterning. At this scale, the symmetry of nano-patterns mimics the orientation of laser polarisation (Fig. 3b). In experiments, laser pulses of $E_p = 3.7 \pm 0.3$ μJ propagate along the $z$-axis; with horizontal, vertical, linear (at 45°) and circular polarisations, with respect to $x$-axis. We exploit $r_0 = 8$ pulses for creating asymmetric arrays of nano-lines, with planar preforms similar to those given in Fig. 2d. The plethora of nano-patterns created in this manner are shown in Fig. 3d. When produced with linear polarisation, their alignment and orientation can be controlled. Further, the feature size in the direction perpendicular to laser polarisation is significantly reduced with linear polarisation (350 nm), compared to that of circular polarisation (800 nm), which otherwise uses identical fabrication parameters. Thus, while symmetric nanofabrication assumes circular polarisation (Fig. 3d-iv), one pushes deeper into the nano-lithography regime with linear polarisation. The unique alignment and orientation control at the nanoscale within the wafer (Fig. 3b) may have exciting implications for future nano-photonics applications. A precursor for such a direction is large volume/area lithographic coverage, experimentally illustrated in Fig. 3e.

Further, by judiciously choosing parameters, such as polarisation, pulse energy, scanning direction, and phase modulation, we achieve record-low feature size for 1D-confined in-chip nano-structures. This ultimately allows sub-diffraction (below 250-nm) lithography (Fig. 4); an exciting new capability implying sub-wavelength nano-photonics within the bulk. Fig 4a shows a set of experiments with $r_0 = 10$ analysing the critical dimension within sub-diffraction regime; as well as the uniformity of lithographic features. The experiments reveal exciting results where ~ 100-nm structures can be created with optical uniformity (Fig. 4a - inset). The SEM image of the first sub-diffraction nano-structures is given in Fig. 4b, indicating $115 \pm 25$ nm lithographic structures, with a sub-wavelength modulation of 800 nm.

Finally, in order to showcase the potential of this unique lithographic capability, we create the first nano-photonics elements deep inside Si (Fig. 5). We start by creating record-efficiency volume-Bragg-gratings (VBG), towards spectral control inside semiconductors. VBGs are based on refractive-index modulation inside transparent media that are patterned over a 3D region[29]. Early Bragg gratings were fabricated in glass, subsequently expanding to other materials and architectures leading to > 90% efficiency and exciting applications[30]. However, analogous components created inside Si are still lacking. In order to remain within the Bragg regime during 3D fabrication, we use the condition, $l > n\Lambda^2 / 2\pi\lambda$, where $l$ is the grating length, $n$ is the average refractive index, $\Lambda$ is the grating period and $\lambda$ is the wavelength in air[29]. Operating in transmission mode, incident light is separated into two distinct orders. The highest efficiency $\eta$, i.e., ratio of power in the diffracted beam to that of total transmission, is obtained for incidence at the Bragg angle[31], $\theta_B$, given by $\sin\theta_B = \lambda/2n\Lambda$. Then, using coupled-wave-theory efficiency for $s$-polarised light can analytically be found as $\eta = \sin^2(\pi l\Delta n/\lambda \cdot \cos(\theta_B))$, where $\Delta n$ is laser-induced index contrast[31].

With these considerations, we start fabricating single-layer Bragg gratings with the feature size of $\xi = 700$ nm, period $\Lambda = 1.5$ μm, and $l = 260$ μm (Fig. 5a). We exploit transverse laser-writing modality, using $r_0 = 7$ and $E_p = 8.7$ μJ (see Methods). A representative VBG operation is given in Fig. 5b, inset. We measured $\eta = \%\ 40 \pm 3.5$ for single-layer gratings, which is already higher than the theoretical limit of $\%\ 33.8$ for a thin sinusoidal phase grating[29]. A further approach for improved efficiency $\eta$ is increasing the grating length $l$. Therefore, we introduced multi-level nano-fabrication inside Si (see Supplementary Note 6). The level-by-level-written VBGs were created over a large area ($3 \times 3$ mm$^2$) with 1.5 hour/level speed, and then characterised with $\lambda = 1.55$ μm linearly polarised light (see Methods). Fig. 5c shows the experimental efficiency values and standard deviations acquired on different positions over grating area, for different lengths, $l$; along with the corresponding theoretical estimates. We observe excellent agreement between experimental and calculated values, assuming an optical index contrast of $|\Delta n| \cong 1.6 \times 10^{-3}$ between laser-written nano-structures and the crystal matrix. The $\Delta n$ value is independently confirmed with quantitative phase microscopy analysis (see Supplementary Note 5). We achieved the highest efficiency of $\%\ 87$ for double-layer gratings of $l = 490$ μm ($l\ /\ \xi = 700$). Further, the angular sensitivity at the Bragg angle is found to be in good agreement with the theoretical estimate (Fig. 5b). The angular bandwidth is measured as $\Delta\theta_{FWHM} = 0.60°$, corresponding to a bandwidth of $\Delta\lambda = 27$ nm.

The preceding results imply that narrow spectral filtering inside Si should be possible. In order to boost angular and, consequently, spectral sensitivity, we fabricated double-level gratings with smaller features (see Supplementary Note 7), achieving sub-micron modulation. The latter is of dimensions $\xi = 350$ nm, $\Lambda = 800$ nm and $l = 430$ μm; created with $r_0 = 7$ and $E_p = 6.6$ μJ. Further, the subsurface levels are seamlessly aligned with no discernible gap leading to increased performance. The zero-order transmitted beam ($I_0$) is used to characterise the VBG (see Methods). Our experimental results successfully illustrate narrower spectral filtering with control on the central wavelength $\lambda_c$, which are in strong agreement with theoretical predictions (Fig. 5d). For instance, while $\theta_{B2} = 16.10°$ corresponds to a bandwidth of $\Delta\lambda = 10$ nm for $\lambda_c = 1545$ nm; $\theta_{B1} = 15.98°$ results in $\Delta\lambda = 8$ nm with $\lambda_c = 1534$ nm (Fig. 5d). In summary, in-chip features and modulations created at the nano-scale enables narrow and tunable spectral filtering capability inside Si. The presented devices constitute the first truly nano-scale functional optical elements that are created completely buried in Si.

In conclusion, we report a methodology for creating controlled nano-structures within the bulk of Si, based on 3D nonlinear laser lithography and a novel seeding effect. We believe the design freedom with nano-lithography in arguably the most important technological material will find exciting applications in electronics and photonics; potentially covering the entire near-/mid-IR regime[20]. The beyond-diffraction-limit features (100 nm) and multi-dimensional confinement imply nano-photonics in Si, such as photonic crystals, meta-surfaces, meta-materials; numerous information processing applications, with significant potential for integration with on-chip systems[17,32-34]. The introduced nano-grating capability is a first step towards this goal, which also constitutes the first multi-layer Si photonics.

## Methods

**Laser nano-lithography system and Bessel beam generation.** A custom-built all-fibre-integrated master oscillator power amplifier system delivers the pulses for volume nano-fabrication. The laser produces pulses in the range of 5 - 30 ns with the maximum power of 9 W, at 150 kHz repetition rate. The central wavelength is $\lambda = 1.55$ μm where Si is transparent. The power and polarisation of the beam are controlled by a set of half-wave-plates (HWP) and/or quarter wave plates (QWP). Additional control is achieved using polarisation beam splitters (PBS), such as to keep laser polarisation along the orientation of liquid crystal molecules over the SLM surface for optimal efficiency. The beam is expanded with a telescope system ($f = 15$ mm, $f' = 35$ mm), before reflected from the SLM (Hamamatsu, liquid-crystal-on-silicon SLM X10468-08, 792 × 600 pixels, 20-μm pixel size). The incidence angle on the SLM is kept lower than 10˚, with a blazed grating superposed to phase mask in order to separate diffraction orders. The hologram surface is rotated to align the +1 order along the optical axis (z-axis), while other diffraction orders are spatially filtered. The reflected beam is imprinted with the information to create a zeroth-order Bessel beam with an axicon-type phase hologram, and is then imaged onto the focusing lens with a 4-f system ($f_1 = 12.5$ cm, $f_2 = 10$ cm). The demagnification of 0.8 is used to completely fill the aperture of the final aspheric lens ($f_3 = 4.5$ mm and NA = 0.55), which is used to focus the beam inside silicon for controlled micro-/nano-fabrication.

For accurate positioning of the beam in Si, the wafer is mounted on a computer-controlled, high-resolution three-dimensional stage (Aerotech, ANT130XY and ANT95-L-Z). For sample preparation, alignment and accurate positioning within the wafer, and selective chemical etching in order to reveal the structures, previously developed protocols are employed[4]. Experiments were performed at a climate controlled laboratory (20 ± 1 °C) and in ambient atmosphere. All transverse and longitudinal laser lithography experiments were performed with 1 - 3 mm/s sample scanning speed.

**Subsurface imaging and material characterisation.** Double-side-polished, 1-mm thick, <100>-cut, p-type (boron-doped, 1-10 Ω.cm) Si samples were used (Siegert Wafer). After laser writing, *in-situ* preliminary analysis is performed with a home-built infrared transmission microscope, comprised of a broad-spectrum halogen lamp and a sensitive complementary metal oxide semiconductor (CMOS) camera (Thorlabs, Quantalux, CS2100M-USB). The resolution of *in-situ* imaging is limited to approximately 1 μm due to the transparency window of Si, thus, for experiments requiring improved resolution, scanning electron microscope (SEM) imaging is employed. In order to image nano-planes, a surface cut is induced with a diamond-tip cutter, which then is propagated through the crystal revealing the cross-sectional subsurface plane (x-z plane). Analysis of the feature size distributions before and upon 40 ± 10 seconds of etching are observed to be consistent. In order to image the nano-lines in Si, mainly SEM is employed, due to the challenges in their *in-situ* analysis. The nano-line images are acquired from the x-y plane, upon polishing with Chemical Mechanical Polishing (CMP), followed by 40 ± 10 s selective chemical etching.

**Simulations for Bessel beam profile.** In order to simulate the Bessel beams created in our setup, we employ scalar diffraction theory for monochromatic waves and use Fresnel diffraction expression[29]. First, the desired phase pattern is digitally loaded onto the SLM (792 × 600 pixels). The incident laser is a Gaussian beam of diameter $2w_0 \cong 4$ mm. The spatially-modulated reflected beam is then imaged onto the aperture of the final focusing lens with a 4-f system ($f_1 = 12.5$ cm, $f_2 = 10$ cm), which is then used in nano-lithography. The complex field distribution in the observation plane after the SLM is calculated with[29],

$$U_2(r,z) = e^{ikz} \cdot \Im^{-1}\left\{ \Im\{U_1(r,0)\}\Im\{e^{-i\frac{\pi}{\lambda z}r^2}\} \right\},$$  (S1)

where $z$ is the distance between the hologram and its projection plane, $\lambda$ is wavelength, $k = 2\pi/\lambda$ is wave-number, $r$ is the polar coordinate with $r^2 = x^2 + y^2$, $U_1(r,0)$ and $U_2(r,z)$ are complex field distributions on hologram and projection planes, respectively. $U_1(r, 0)$ is expressed as $U_1(r, 0) = U_0 \emptyset_{\text{SLM}}(r)$, where $U_0$ is the incident field amplitude and $\emptyset_{\text{SLM}}(r)$ is the phase pattern digitally loaded to SLM. In our case, different phase distributions are analogous to employing axicons of different cone angles. The demagnification of the system is included in simulations by employing a demagnification factor of $f_2/f_1$= 0.8 for pixel size and beam diameter. Since the 4-$f$ system translates the SLM phase distribution to the final focusing lens ($f_3$ = 4.5 mm and NA = 0.55), the phase profile of lens $\emptyset_{lens}(r) = \exp(-i\pi r^2/\lambda f)$ is also multiplied with $U_1(r, 0)$. The expression in the diffraction equation then takes the form, $U_1(r, 0) = U_0 \emptyset_{\text{SLM}}(r)\emptyset_{\text{lens}}(r)$. In experiments, we used $\phi_{\text{SLM}}(r) = \exp(+i2\pi\, r/r_0)$ corresponding to a diverging axicon for positive phase, where decreasing the $r_0$ parameter is equivalent to increasing axicon cone angle. The simulations for the field intensity after the focusing lens and with various $r_0$ values are shown in Fig. 2 and Supplementary Information, respectively. The simulations, employing 2D Fast Fourier Transform (FFT) algorithm, are done in MATLAB. Finally, the intensity distribution is calculated from $I_2(x, y) = |U_2(x, y)|^2$.

**Volume Bragg Grating fabrication and optical characterisation.** Single-layer gratings were written with transverse laser-writing modality, using a scanning speed of 1 mm/s, repetition rate of 150 kHz, $r_0$ = 7, and $E_p$ = 6.6 - 8.7 µJ. The polarisation of the laser remains parallel to the sample scanning direction during fabrication. After laser writing, VBGs were imaged with an IR transmission microscope verifying their morphology, followed by cross-sectional imaging with SEM for more detailed analysis. Multi-layer VBGs are fabricated with the same set of lithography parameters, however, writing the deepest layer first and consecutively moving closer towards the beam entrance surface for each layer. All gratings were created with a surface area of $3 \times 3$ mm$^2$ in the plane perpendicular to $l$, with 1.5 hour/level fabrication speed. Efficiency tests were performed with a diode laser of $\lambda$ = 1.5 µm (Thorlabs, FPL1009S), where Si is transparent. The diode laser is on the $x$-$z$ plane, and is of $s$-polarisation ($y$-axis) before the sample. The diffracted wave is imaged with an Indium Gallium Arsenide (InGaAs, Artray Artcam-990SWIR) camera. In order to take into account any losses due to scattering or reflection, efficiency definition is used as the ratio of power in the +1 diffraction order ($I_{+1}$) to that of total transmitted light ($I_0 + I_{+1}$).

In order to analyse the spectral response of VBGs, a $c$-band, $p$-polarised Amplified Spontaneous Emission (ASE) source covering the spectral range of 1520 nm-1560 nm is used. The zero-order transmitted beam ($I_0$) is coupled to a collimator and directed to an optical spectrum analyser (OSA, Exfo FTBX-5235). The transmission spectrum is then normalised to the spectrum of light passing outside the grating area. The spectral response analysis is repeated for multiple incidence angles. For analytical modelling of efficiency, with its angular and spectral response, we use the general diffraction efficiency equation[31], $\eta = \sin^2(\Phi^2 + \varXi^2)^{1/2}/(1 + \varXi^2/\Phi^2)$, where $\Phi$ is the phase term determining the highest efficiency depending on polarisation and duty cycle, and $\varXi$ is the dephasing parameter which explains deviations from the Bragg condition by detuning $\theta_B$ or $\lambda_c$.

**Data availability.** The data that support the plots within this paper and other findings of the study are available from the corresponding author upon reasonable request.

# Figure Captions

**Figure 1 | Concept of seeding for nano-lithography inside Si. a,** Spatially-modulated nanosecond laser pulses are used to create non-diffracting beams inside Si. These are exploited to induce planar nano-patterns deep inside silicon (purple coloured region), without altering the wafer above or below the modifications. Inset: Scanning Electron Microscope (SEM) image of structures with beyond-diffraction-limit features written inside Si. **b,** Using these preformed structures, *i.e.*, in-chip preforms, one can nonlocally seed 2D-confined patterns or nano-lines buried inside Si (yellow region). While the lateral range for seeding is ~ 3 μm, longitudinal writing modality allows the nano-lines to be extended throughout the entire wafer, including to regions where the seed is absent. Similarly, the nano-lines can be fabricated laterally, creating large-area-covering volumetric nano-patterns. Inset: SEM image of the cross-section of large-area covering nano-lines created with non-local seeding. The alignment and symmetry of the nano-lines can be controlled with the laser polarisation.

**Figure 2 | Laser nano-lithography inside Si. a,** Schematic showing Bessel beam generated with diverging-axicon and lens doublet. $L_1$ and $L_2$ serve as 4-*f* system that translates the SLM output to aspheric lens ($L_3$) with a demagnification factor of 0.8. **b,** Beam characterisation in air for $r_0$ =10. **(i)** Comparison of simulated and experimental intensity profiles in the transverse plane at maximum intensity plane. Concentric rings are visible in the infrared camera (InGaAs) inset image. **(ii)** Experimental intensity profile recorded on *x-z* plane. **c,** 1D- and 2D-confined lithography. **(i)** Scanning electron microscope (SEM) image of the wafer cross-section revealing nano-planes created inside Si. **(ii)** Colour-coded schematic for subsurface planes. The blue cage represents the cross-sectional plane for laser-written structures of (i). The purple cage indicates the cross-sectional plane given in (d-i). The red cage captures an isolated nano-line indicated in (d-iii). **d,** SEM images at different depths. **(i)** SEM image with purple border is acquired at a deeper plane, capturing both the seed and nano-line. The seed is created with $r_0 = 10$ while the nano-line is created with $r_0 = 6$. **(ii)** The image with black border shows nano-lines do not form in the absence of a seed. **(iii)** SEM image with red border shows an individual nano-line. 40 seconds etching was used to reveal the structures.

**Figure 3 | Polarisation-dependent nano-fabrication. a,** Subsurface nano-lithography with 1D confinement, as a function of $r_0$ and $E_p$. **b,** Schematic of polarisation-dependent fabrication with 2D-confinement. Blue sections indicate laser-written nano-patterns with controllable orientation inside Si. **c,** SEM image of highly-reproducible, uniform nano-planes created with $r_0$ =10. The structures have a thickness of $250 \pm 30$ nm close to the diffraction limit. The laser polarisation is parallel to the scanning direction. **d,** Architecture of 2D-confined nano-patterns created with various laser polarisations. The thickness along the short axis is 350 nm for linearly-polarised laser writing. The arrays are created with longitudinal-writing modality, exploiting a Bessel beam of $r_0 = 8$ and $E_p = 3.7 \pm 0.3$ μJ. SEM scale bar = 1 μm. **e,** SEM image of large-area-covering nano-array created in Si ($r_0 = 8$ with $E_p = 3.7$ μJ). All samples are polished and etched for 40 seconds to reveal the structures.

**Figure 4 | Sub-diffraction fabrication regime. a,** Plot of mean feature size vs. pulse energy for $r_0 = 10$, illustrating the wide range of control that can be exerted in both nano- and sub-diffraction regimes for 1D confinement. The histogram in the inset shows the size distribution of laser-written structures for $E_p = 3 \pm 0.3$ μJ. **b,** The SEM image indicating sub-diffraction fabrication with nano-planes with $\xi = 115 \pm 25$ nm and sub-wavelength modulation of 800 nm, for $r_0 = 10$ and $E_p = 3 \pm 0.3$ μJ. Chemical etching was performed for 40 seconds prior to SEM imaging.

**Figure 5 | In-chip nano-photonics. a,** Schematic of VBG buried in Si. Nano-planes are created with feature size $\xi$, pitch $\Lambda$, and length $l$, controlled with multi-level nano-fabrication. Incident light, $I_{in}$, is separated into two orders over *x-z* plane. **b,** Angular sensitivity for two-level VBG with $\xi = 700$ nm, $\Lambda = 1.5$ μm, $l = 490$ μm, recorded with *s*-polarised single-mode laser of $\lambda = 1550$ nm. The angular bandwidth is measured as $\Delta\theta_{FWHM} = 0.60°$, corresponding to a bandwidth of $\Delta\lambda = 27$ nm. Theoretical prediction is given with Kogelnik theory for thick gratings. Inset: Diffraction patterns imaged with InGaAs camera. **c,** Diffraction efficiencies measured at the Bragg angle for multi-level gratings of $\xi = 700$ nm, $\Lambda = 1.5$ μm, and various grating lengths, $l$. Data points correspond to one- to four-level writing. Error bars indicate standard deviation. Black curve represents corresponding theoretical calculation. Maximum efficiency of %87 is measured for two-level VBG, $l = 490$ μm. **d,** Spectral sensitivity of VBG created with $\xi = 350$ nm, $\Lambda = 800$ nm, and $l = 430$ μm. Spectrum for $I_0$ is measured at the Bragg angle $\theta_{B1}$ = 15.98° resulting in $\Delta\lambda = 8$ nm and $\lambda_c = 1534$ nm; while $\theta_{B2} = 16.10°$ corresponds to a bandwidth $\Delta\lambda = 10$ nm, $\lambda_c = 1545$ nm.

# Acknowledgments

This work was supported partially by the TÜBITAK under projects 219M274, 121F387 and by the Turkish Academy of Sciences, TÜBA-GEBIP Award. The authors also thank Dr. F. Ömer Ilday for inspiration.

# Author contributions

R.A.S. and O.T. designed the research and interpreted the results. Experiments were performed by R.A.S., A.I., A.S. and O.T. Simulations were performed by R.A.S, and phase microscopy analysis is performed by A.S.

**Figure 1**

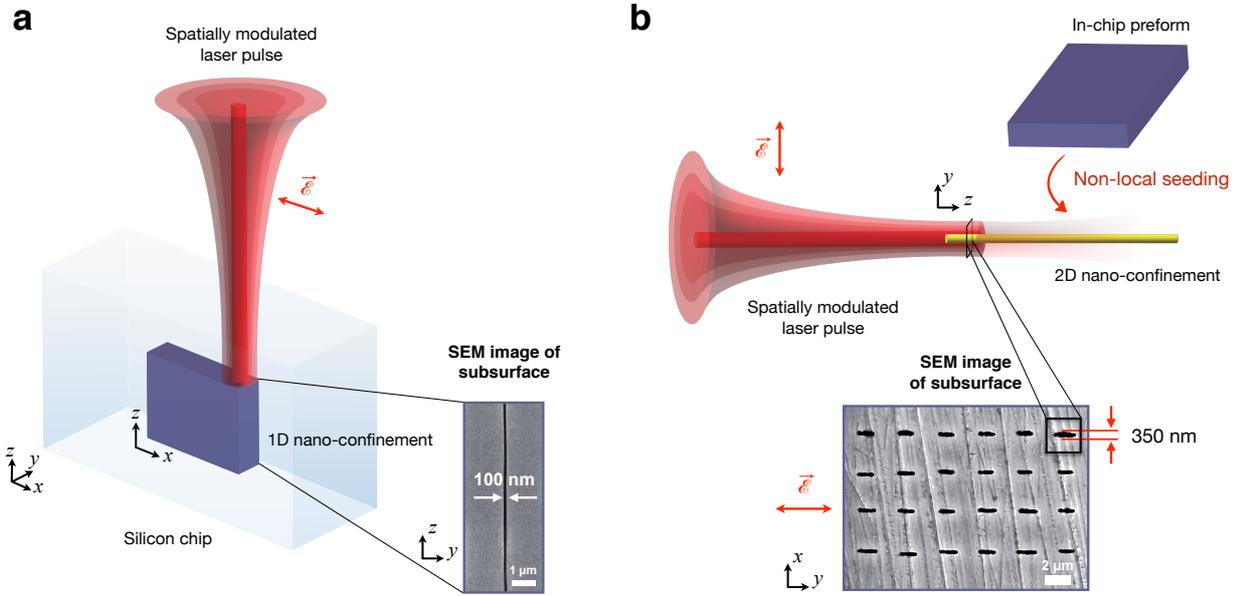

**a**

Spatially modulated
laser pulse

$\vec{e}$

1D nano-confinement

**SEM image of
subsurface**

100 nm

1 μm

Silicon chip

**b**

In-chip preform

Non-local seeding

$\vec{e}$

2D nano-confinement

Spatially modulated
laser pulse

**SEM image
of subsurface**

350 nm

$\vec{e}$

2 μm

**Figure 2**

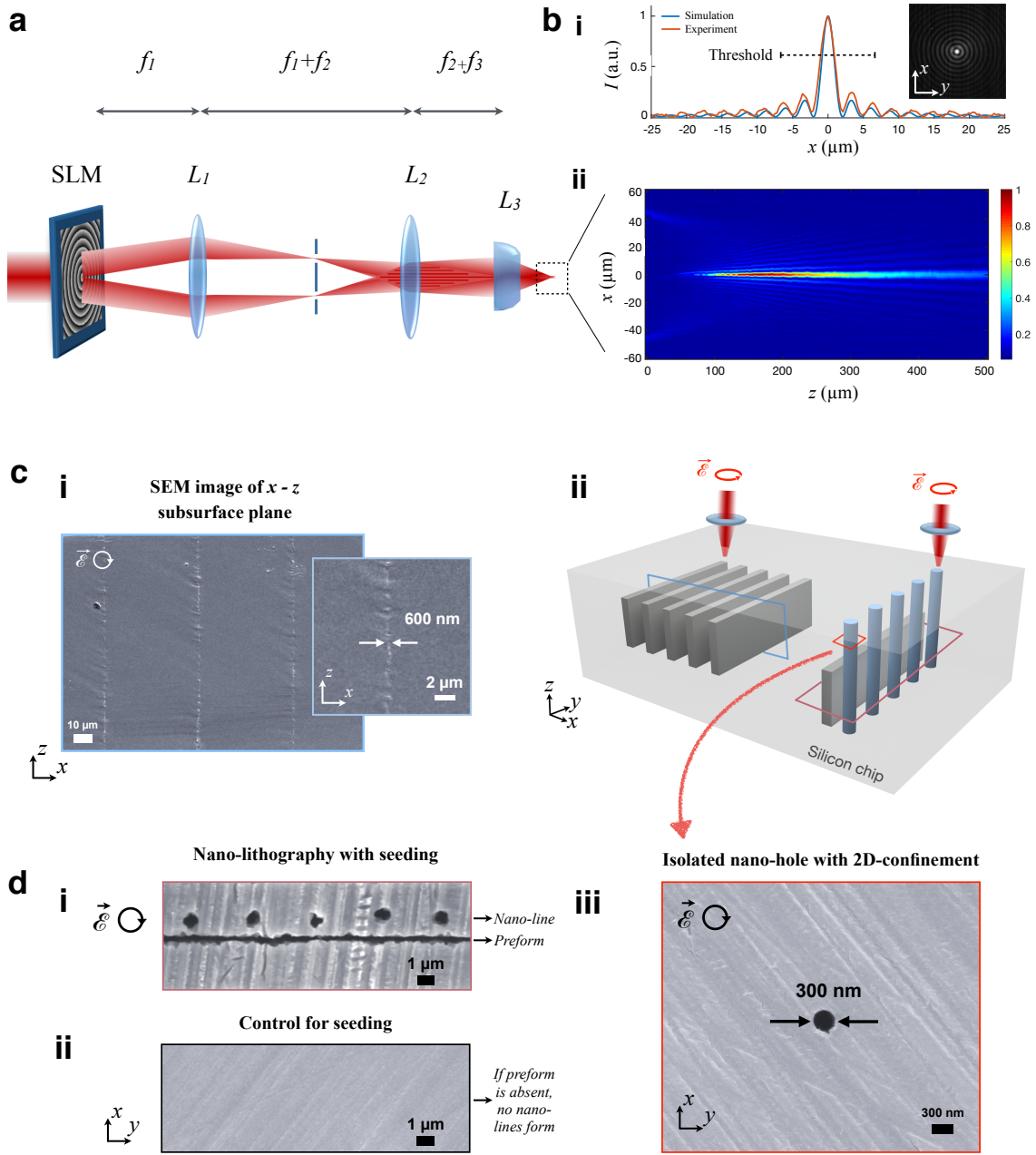

**Figure 3**

**a**

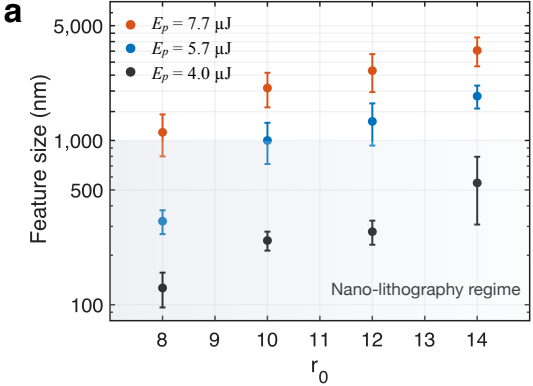

**b**

Polarisation controlled nano-lithography

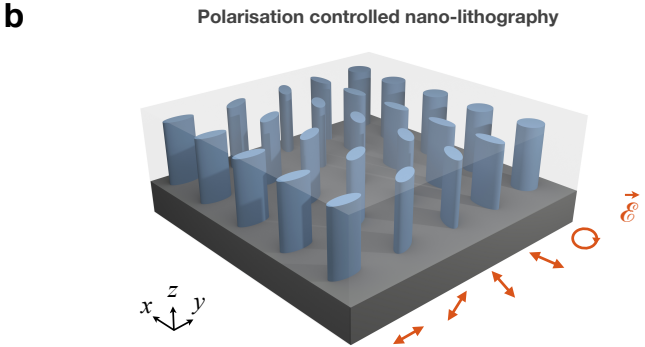

**c**

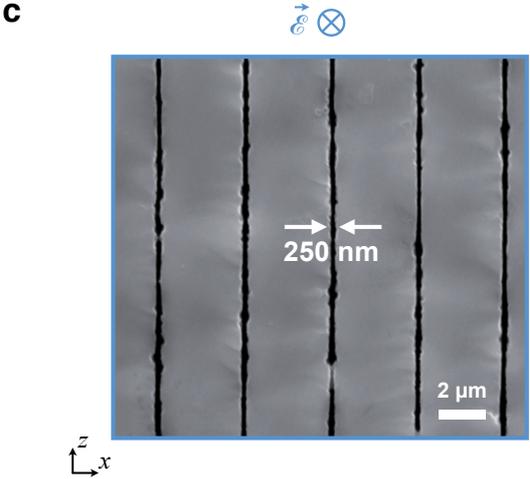

**d**

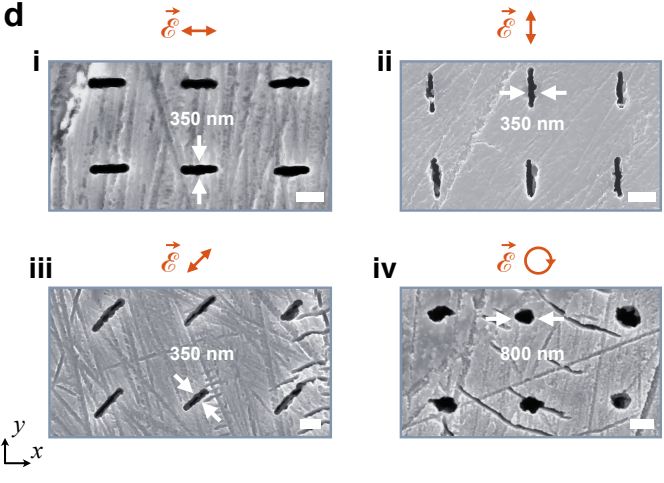

**e**

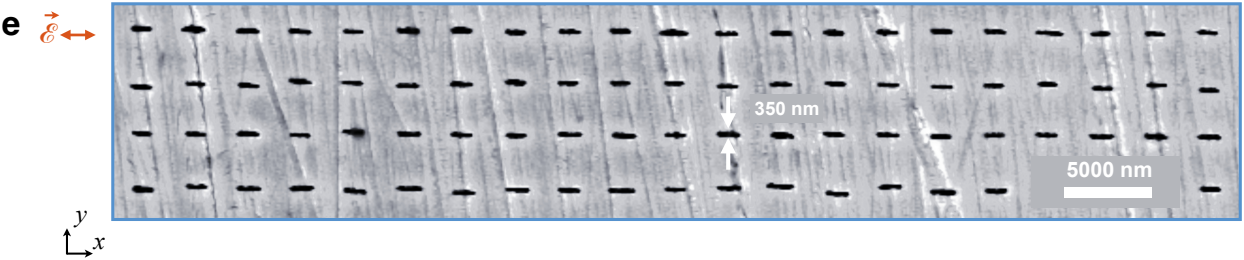

**Figure 4**

**a**

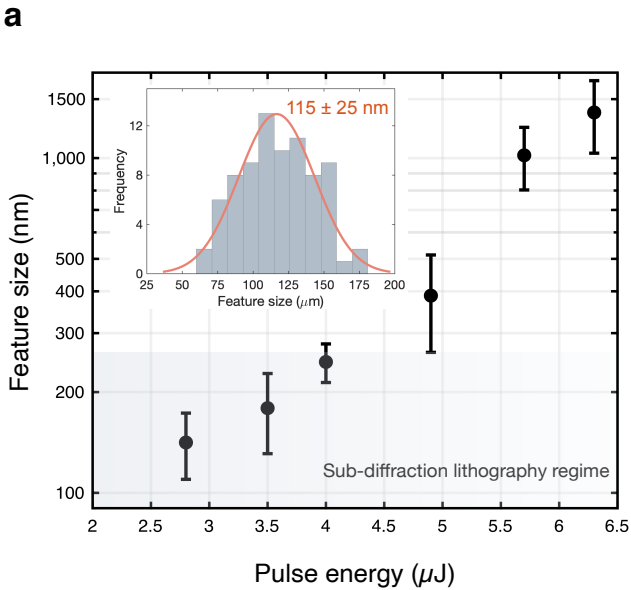

**b**

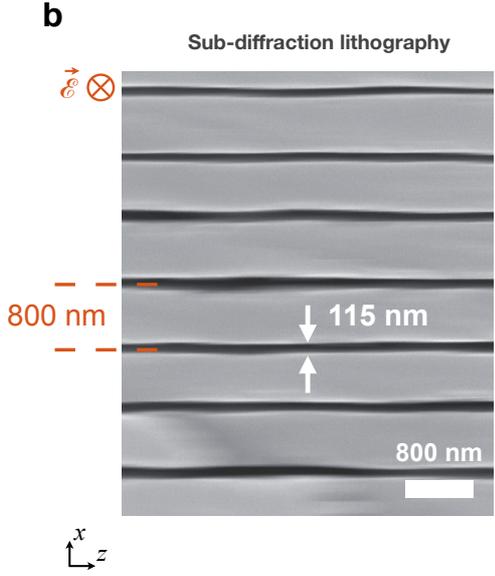

Sub-diffraction lithography

# Figure 5

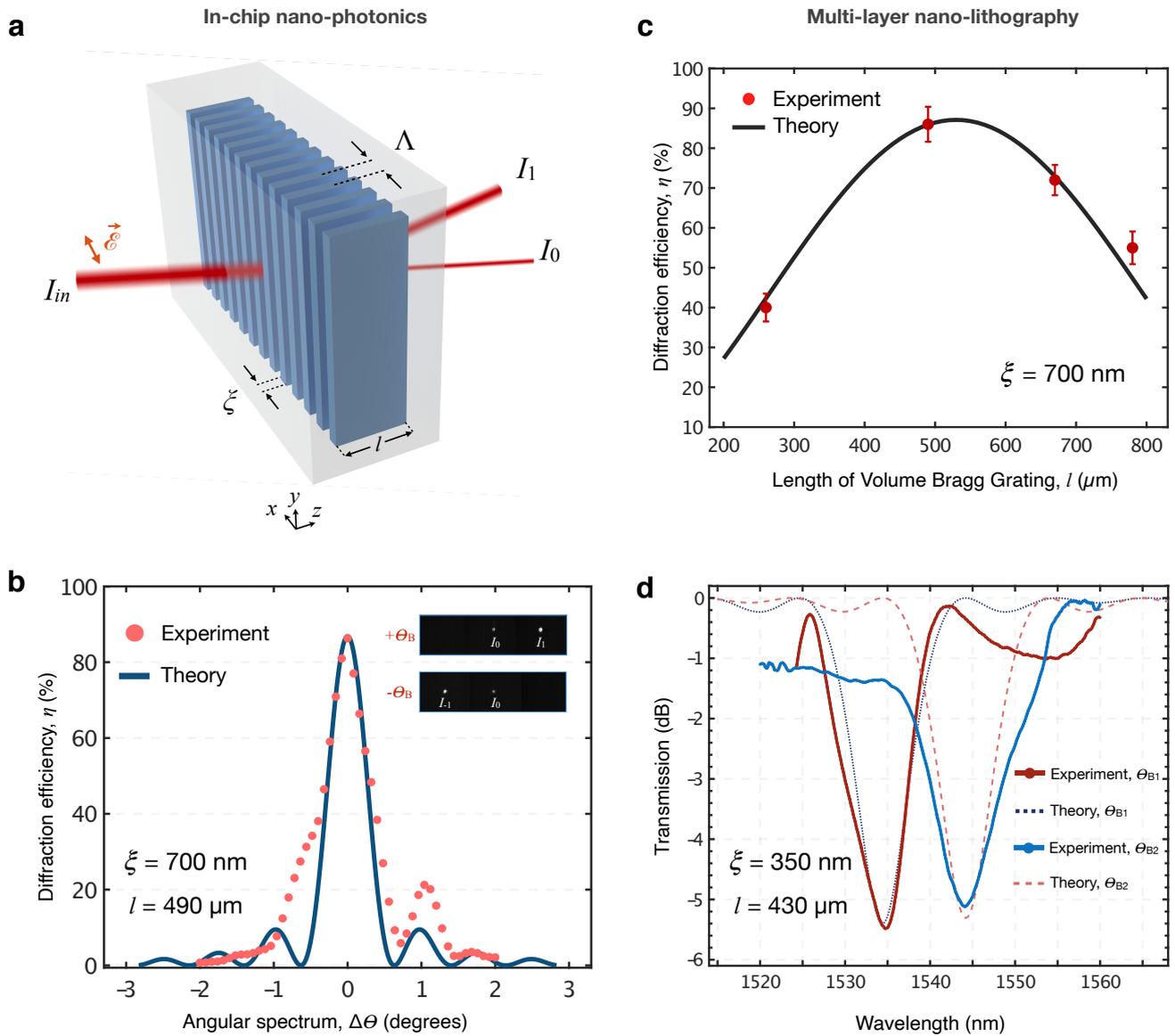

**a** In-chip nano-photonics

**c** Multi-layer nano-lithography

**b**

**d**